\documentclass[12pt]{iopart}

\usepackage{cite}
\usepackage{graphicx}
\newcommand{\beq}{\begin{equation}}
\newcommand{\eeq}{\end{equation}}
\newcommand{\beqa}{\begin{eqnarray}}
\newcommand{\eeqa}{\end{eqnarray}}
\begin{document}


\title[Obliquely propagating solitary waves in magnetized kappa-distributed plasmas]
{Oblique propagation of arbitrary amplitude electron acoustic solitary waves in magnetized kappa-distributed plasmas}
\author{S. Sultana$^1$, I. Kourakis$^1$ and M. A. Hellberg$^2$}

\address{$^1$Centre for Plasma Physics, Department of Physics and
Astronomy, Queen's University Belfast, BT7 1NN Northern Ireland,
UK}

\address{$^2$School of Chemistry and Physics, University of KwaZulu-Natal,
    Private Bag X54001, Durban 4000, South Africa}

\ead{basharminbu@gmail.com, IoannisKourakisSci@gmail.com and hellberg@ukzn.ac.za}
\begin{abstract}
The linear and nonlinear properties of large amplitude electron-acoustic waves are investigated in a magnetized plasma comprising two distinct electron populations (hot and cold) and immobile ions.
The hot electrons are assumed to be in a non-Maxwellian state, characterized by an excess of superthermal particles, here modelled by a kappa-type long-tailed distribution function.
Waves are assumed to propagate obliquely to the ambient magnetic field.
Two types of electrostatic modes are shown to exist in the linear regime, and their properties are briefly analyzed.
A nonlinear pseudopotential type analysis reveals the existence of large amplitude electrostatic solitary waves and allows for an investigation of their propagation characteristics and existence domain, in terms of the soliton speed (Mach number).
The effects of the key plasma configuration parameters, namely, the superthermality index  and the cold electron density, on the soliton characteristics and existence domain, are studied.
 The role of obliqueness and magnetic field are discussed.

\end{abstract}

\maketitle

\section{Introduction}

 A high frequency electrostatic mode exists in plasmas containing two distinct temperature electrons (here referred to as ``hot" and ``cold" electrons) and ions \cite{watanabe1977,tokar1984,gary1985,mace1990}.
 For these electron-acoustic waves (EAWs) to be sustained, the restoring force comes from the pressure of the hot electron component and the inertia is provided by the cold electrons.
  The \textbf{electron-acoustic (EA) wave} (EAW) frequency is much higher than the ion plasma frequency, therefore the ion species may be considered as a neutralizing background, which does not influence the dynamics. The phase speed $v_{ph}$ of EAWs is much lower than the hot electron thermal speed $v_{th,h}$ and much higher than the cold electrons thermal speed $v_{th,c}$, i.e., $v_{th,c}\ll v_{ph}\ll v_{th,h}$, in order to avoid wave damping due to resonance with either the hot or the cold electrons. Here, the indices $h$ and $c$ refer to the hot and the cold electron, respectively. Furthermore, since  $v_{th,h/c} \sim T_{h/c}^{1/2}$, the hot electron temperature  $T_{h}$ should be much greater than $T_c$;
 \emph{ab initio} kinetic theory studies have shown that EAW can exist only when $T_{h}/T_{c}\gg 10$ and when the cold electron component corresponds to $20\%$ to $80\%$ of the total electron population \cite{tokar1984,gary1985,mace1990,Amery,BHJGR}.

Many theoretical investigations have been carried out since the original work of Watanabe~\cite{watanabe1977}, where the existence of EAWs was first predicted~\cite{tokar1984,gary1985,mace1990,Amery,BHJGR,hellberg2000}. The occurrence of EAWs is now  confirmed by FAST satellite observations associated with  auroral density cavities~\cite{pottelette1999} and also in the mid-altitude auroral zone~\cite{ergun1998} (the interested reader is also referred to the references cited in Refs.~\cite{pottelette1999} and~\cite{ergun1998}). Moreover, laser plasma experiments~\cite{montgomery2001} have traced the signature of EAWs. The nonlinear propagation of EAWs has been investigated by many researchers, both in unmagnetized~\cite{SSIK1011,sultana2011,danehkar2011,mamun_unmag2002,singh2004} and in magnetized plasmas~\cite{kourakis2004,mace2001,mamun2002}.
The EA solitary wave characteristics in unmagnetized plasma were studied in Ref.~\cite{mace1991}, where the existence of negative potential electrostatic structures associated with cold electron density compression regions was established. On the other hand, magnetized plasma have been shown to sustain EA solitary waves, in the form of pulse-shaped small-amplitude solitary waves in the long-wavelength limit~\cite{mace2001,mamun2002}. The evolution of modulated wavepacket propagation into envelope solitons due to oblique perturbations of the wave amplitude has also been demonstrated~\cite{kourakis2004} in a Maxwellian plasma. Magnetized as well as unmagnetized ions with Maxwellian hot electrons were considered in Ref.~\cite{mace2001}, where the existence of negative polarity solitary waves was shown, when a drift was present.

The properties of large amplitude EA solitary waves in ummagnetized nonthermal plasmas have been investigated in
Ref.~\cite{singh2004}, where nonthermality was introduced via a Cairns type non-Maxwellian distribution of the hot electrons which, interestingly enough, account for negative potential solitary structures. An analogous study has been presented in Ref.~\cite{danehkar2011}, where solitary wave occurrence in unmagnetized plasma was considered via a pseudopotential method, with kappa-distributed hot electrons. The existence of positive polarity EA solitons was predicted~\cite{mamun2002} in a nonthermal plasma by introducing a vortex type hot electron distribution in a magnetized plasma.

Abundant space plasma observations~\cite{vocks2003,gloeckler2006,vocks2008}
and experiments~\cite{hellberg2000,preische1996,yagi1997,magni2005} have reported the presence of highly energetic particles. The observed velocity distribution can be fitted by a $\kappa-$type (generalized Lorentzian)
function~\cite{vasyliunas1968,summers1991,hellberg2000,hellberg2009}, where $\kappa$ defines the strength of superthermality; the lower the value of $\kappa$, the higher the degree of excess superthermality, or in other words, the deviation from the Maxwellian behavior.
Refs.~\cite{Amery,hellberg2000,BHJGR}, on the one hand, and~\cite{danehkar2011}, on the other,  have studied the effects of $\kappa$-distributed hot electrons on linear and nonlinear electron acoustic waves in unmagnetized plasmas, respectively.

Our main aim in this article is to investigate the characteristics of large amplitude electrostatic solitary waves of  electron-acoustic type in magnetized plasmas characterized by the presence of an excess population in the superthermal hot electron component. As a hypothesis, the hot electrons are considered to be $\kappa-$distributed.
The manuscript at hand is arranged in the following manner. The basic formalism, relying on a multi-fluid model, is presented in Section~\ref{secModel}. The characteristics of the linear waves are analyzed in Sec.~\ref{linearanalysis}. A fully nonlinear pseudopotential-type  approach is presented in Sec.~\ref{secSagdeev} to study  arbitrary amplitude EA solitary waves. The existence domains for EAW propagation are delineated in Sec.~\ref{speeddomain} and their parametric dependence on the relevant plasma parameters is analyzed. The soliton characteristics are investigated in
Sec.~\ref{EAcharac}. Our findings are summarized in the concluding Section~\ref{EAsagconc}.

\section{Basic formalism \label{secModel}}

We consider a three component plasma model consisting of two electron components (referred to as ``cold'' and ``hot'') and ions.
The ambient magnetic field $\mathbf{B_0} (= B_0 \widehat{z})$ is assumed stationary, pointing along the $z-$axis.
The direction of wave propagation, at an angle $\theta$ to $\bf{B_0}$, together with the magnetic field  direction,  defines the $x-z$ plane.
We also introduce the direction cosines $\alpha$ and $\gamma = \cos\theta$ with respect to the $x-$ and $z-axis$, respectively.

The inertia of the cold electrons is retained via a fluid model, that satisfies  $v_{th,c}\ll v_{ph}\ll v_{th,h}$.
The (inertialess) hot electrons have a non-Maxwellian character, described by a $\kappa-$type velocity distribution,
that leads to a hot electron number density  of the form~\cite{hellberg2009}
\begin{equation}
n_{h}=n_{h0} \left[1-\frac{ e \Phi}{(\kappa-\frac{3}{2}) k_B T_h}\right]^{-\kappa +1/2}\ , \label{nek}
\end{equation}
where $n_{h0}$ represents the hot electron number density at equilibrium, $k_B$ is the Boltzmann constant and $T_h$ is the characteristic ``temperature'' of the hot electrons~\cite{hellberg2009}.
On the timescale of interest, the ions may be considered as stationary.

The dynamics of the cold electron component are described by the following (normalized) equations
\begin{eqnarray}
\frac{\partial n}{\partial t} &+& \nabla \cdot (n\,\mathbf{u}) = 0\ ,\label{eo1}\\
\frac{\partial \mathbf{u}}{\partial t} &+& (\mathbf{u}\cdot \nabla)\mathbf{u} - \nabla \phi - \Omega_{c}(\mathbf{u}\times\widehat{z}) = 0\ ,\label{eo2}\\
\nabla^{2} \phi &=& \beta(n-1)+\biggr(1-\frac{\phi}{\kappa-\frac{3}{2}}\biggr)^{-\kappa+\frac{1}{2}}-1\ , \label{easagpois1}
\end{eqnarray}
where the cold electron number density $n_c$, velocity $u_c$, and electrostatic potential $\Phi$ are normalized as $n\equiv n_c/n_{c0}$,
$u = u_{c}/v_s$  and $\phi \equiv  \Phi/\Phi_0$, respectively.
Here, $\Omega_{c}=e\mathbf{B_0}/({m_e}\omega_{ph}) =B_{0}/\sqrt{4\pi n_{h0} m_{e}}$ is the (electron) cyclotron frequency normalized to the (hot electron) plasma frequency $\omega_{ph} = (4\pi n_{h0} e^2/m_{e})^{1/2}$.
Space and time variables are scaled by the hot electron Debye length
$\lambda_{Dh}= \left( k_B T_{h}/4 \pi n_{h0} e^2 \right)^{1/2}$,
and the inverse hot electron plasma  frequency, $\omega_{ph}^{-1}$, respectively.
Finally, the electric potential $\Phi$ is scaled as $\phi = \Phi/\Phi_0 = e \Phi/k_B T_{h}$.
The characteristic speed scale is the hot electron thermal speed $v_s\equiv (k_B T_h/m_e)^{1/2}$. We have defined the parameter $\beta=n_{c0}/n_{h0}$, denoting the cold-to-hot electron population ratio.

\section{Linear waves \label{linearanalysis}}

In the small amplitude limit, we may consider perturbations varying as $e^{i (\mathbf{k} \cdot \mathbf{r}-\omega t)}$, and thus derive a dispersion relation, relating the wave frequency $\omega$ to the wave number
$k$. By linearizing the dimensionless evolution equations (\ref{eo1}) - (\ref{easagpois1}), we obtain the dispersion relation in the form
\begin{eqnarray}
\omega_{C,A}^{2}=\frac{1}{2}\biggr(\Omega_{c}^{2}+\tilde{\omega}^{2}\biggr)\left[1 \pm \left(1-\frac{4\Omega_{c}^{2}\,\tilde{\omega}^{2}\cos^{2}\theta}{(\Omega_{c}^{2}+\tilde{\omega}^{2})^2}\right)^{1/2}\right]\ , \label{easagdis1}
\end{eqnarray}
where
\begin{equation}
\tilde{\omega}^{2}=\frac{\beta k^{2}}{k^{2}+c_{\kappa}}\ ,\label{freq}
\end{equation}
with $c_{\kappa}=(\kappa-\frac{1}{2})/(\kappa-\frac{3}{2})$.
 Note that Eq. (\ref{freq}) essentially provides the wave frequency in an unmagnetised
 plasma~\cite{kourakis2004,sultana2011}; cf. the results in Ref.~\cite{danehkar2011}.
 The parameter $\beta$ appears due to the fact that only a fraction of the electrons are taking part in the shielding effect.

The plasma model described above therefore supports two linear modes with frequencies $\omega_{C}$ and $\omega_{A}$.
The upper/lower sign in~(\ref{easagdis1}) corresponds to the index C/A, respectively, on the left-hand side of Eq.~(\ref{easagdis1}).
Upon restoring dimensions, it is straightforward to show that in the Maxwellian limit $\kappa \rightarrow \infty$,  the dispersion relation~(\ref{easagdis1}) is in  agreement with the dispersion relation given in
Ref.~\cite{mace1993}.
The wave with frequency $\omega_{C}$ (fast mode) represents electron cyclotron-like (EC) waves (ECWs), that are modified by acoustic effects, while the wave with frequency $\omega_{A}$ (slow mode) represents obliquely propagating EAWs.
Both modes are strongly dependent on the superthermal character of the plasma, and so is the charge screening mechanism, as denoted by the appearance of the factor $c_{\kappa}$ in the denominator in~(\ref{freq}). The phase speed of the obliquely propagating EA mode [lower sign in Eq.~(\ref{easagdis1})] in the limit $k^{2}\ll c_{\kappa}$ and $\tilde{\omega}^{2}\ll \Omega_{c}^{2}$ reads:
\begin{equation}
v_{ph}^{(-)} = \cos\theta\sqrt{\beta/c_{\kappa}}\, . \label{EAphasemag}
\end{equation}

In the vanishing magnetic field limit, the dispersion relation~(\ref{easagdis1}) trivially recovers unmagnetized high-frequency EAWs [given by relation~(\ref{freq})], which agrees with Refs.~\cite{sultana2011}
and~\cite{kourakis2004}.
Naturally, the same dispersion relation (\ref{freq}) is recovered upon setting $\theta \rightarrow 0$
in~(\ref{easagdis1}), for parallel propagation (recall that the Lorentz force has  no component in the direction of the magnetic field).
In this case, in the long wavelength limit, i.e., for $k^{2}\ll c_{\kappa}$, the wave frequency becomes $\omega_{C} \to \,\sqrt{\beta/c_{\kappa}}$, so that the phase velocity is given by
\begin{equation}
v_{ph}=\sqrt{\beta/c_{\kappa}} \, ,  \label{vph}
\end{equation}
in agreement with previous results in Ref. \cite{sultana2011} (considering $c_{\kappa} = c_{1}$ therein) and also with Ref. \cite{kourakis2004} in the Maxwellian case (recovered upon setting $c_{\kappa}=1$).
A detailed analysis of this mode can be found in Refs.~\cite{sultana2011} and~\cite{kourakis2004}.
As we will see later in Section~\ref{speeddomain}, the phase speed $v_{ph}$ provides the velocity threshold for (superacoustic) EA solitary excitations to exist, and thus corresponds to the real sound (acoustic) speed in the plasma, as will be discussed below.
It should be added, for rigor, that we need to consider a phase speed much lower than the hot electron thermal speed,
in order to validate the fluid description~\cite{watanabe1977} for cold electrons (i.e., $v_{th,c}\ll v_{ph}\ll v_{th,h}$). The requirement ensuring the validity of our model thus translates as:
\begin{equation}
\frac{\beta}{c_{\kappa}}\ll 1 \qquad \Rightarrow \qquad \beta \ll \frac{\kappa-1/2}{\kappa-3/2}\ . \label{relnbetakappa}
\end{equation}
We therefore consider values of the cold-to-hot electron density ratio $\beta$ obeying Eq. (\ref{relnbetakappa}),
for the respective values of the superthermality parameter $\kappa$, both in our linear and nonlinear analyses. For example, for $\kappa = 3$, $5$ and $100$, the upper limit~(\ref{relnbetakappa}) implies that $\beta \ll 1.67$, $1.29$ and $1$, respectively.

It may be instructive to analyze from first principles the dependence of electron-acoustic waves on various relevant plasma model parameters.
Linear wave characteristics depend on: \\
 --- electron superthermality (quantified via $c_{\kappa}$):  the lower the $\kappa$ value the stronger the superthermality (the Maxwellian case is
 recovered for $c_{\kappa}\rightarrow 1$, i.e., $\kappa \rightarrow\infty$);\\
 --- the cold-to-hot electron concentration via $\beta$;
in fact, $\beta$ should be between $0.25$ and $4$  approximately, in order for Landau damping to be
minimized~\cite{mace1990,Amery,BHJGR}, the upper limit being further reduced by expression~(\ref{relnbetakappa});\\
---  the magnetic field, via the normalized hot electron cyclotron frequency $\Omega_c$; and
\\
--- obliqueness (via $\theta$).

\begin{figure}[!h]
 \begin{center}
 \includegraphics[width=8cm]{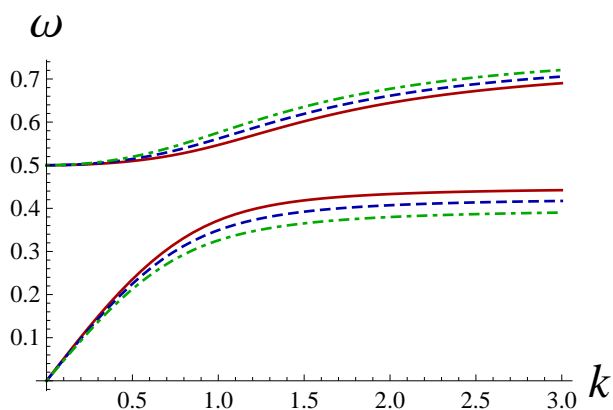}
 \end{center}
 \caption{(Color online) \emph{Effect of propagation angle.}The linear dispersion relation (Eq.~\ref{easagdis1}) is depicted for obliquely propagating electrostatic waves in a magnetized plasma, for $\beta=0.5,\,\kappa=3$ and $\Omega_{c}=0.5$.
 The upper (three) curves for $\omega_C$ and the lower (three) curves for $\omega_A$ correspond to the fast (electron cyclotron-like) and slow (electron acoustic) mode, respectively.
 The curves have been calculated for different angles  with respect to the external magnetic field ${B_0}$: $\theta=20^{\circ}$ (solid curve);  $\theta=25^{\circ}$ (dashed curve);  $\theta=30^{\circ}$ (dot-dashed curve). }
 \label{Fig.1}
\end{figure}

\emph{Effect of oblique wave propagation.} Fig.~\ref{Fig.1} shows the effect of propagation angle  on the linear properties of the electrostatic modes found above: the upper branch corresponds to obliquely propagating electron cyclotron-like waves, while the lower branch is for obliquely propagating EAWs.
Increasing the angle with respect to  the magnetic field leads to a decrease in the frequency $\omega$ of the lower (EA) mode, and a frequency increase in the  electron cyclotron mode, along with the analogous modification in the phase speeds for both modes.
This is related, effectively, to the fact that the phase speed of the slow mode varies as $\cos \theta$, while the restoring force, and hence the phase speed, for the cyclotron modes varies as $\sin\theta$.
Interestingly, a similar trend, increasing the frequency gap between the two modes,  was found for oblique ion-acoustic waves in Ref.~\cite{sultana2010} (see Fig. 3 therein).

\begin{figure}[!h]
 \begin{center}
 \includegraphics[width=8cm]{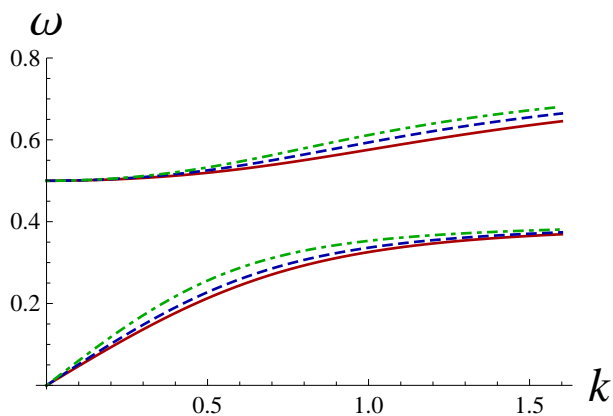}
 \end{center}
 \caption{(Color online) \emph{Effect of superthermality.} The linear dispersion relation for electrostatic waves propagating at an angle $\theta=30^{\circ}$ with an external magnetic field, for $\beta=0.5$, $\Omega_{c}=0.5$.
 The upper (three) curves and the lower (three) curves depict $\omega_C$ and $\omega_A$, respectively. Different values of $\kappa$ have been considered:
 $\kappa=3$ (solid curve); $\kappa=5$  (dashed curve); and $\kappa=100$ (dot-dashed curve).}
 \label{Fig.2}
\end{figure}

\emph{Effect of superthermality.} The effects of superthermality  on
obliquely propagating EA and electron cyclotron waves are depicted in Fig.~\ref{Fig.2}.

It is seen that the more strongly non-Maxwellian distributions (lower $\kappa$, increased excess superthermal particles) lead to both modes having lower frequency and phase velocity.

\begin{figure}[!h]
 \begin{center}
 \includegraphics[width=8cm]{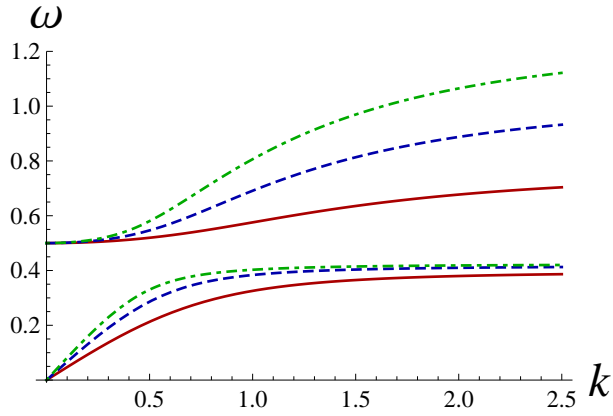}
 \end{center}
 \caption{(Color online) \emph{Effect of the cold-to-hot electron density ratio $\beta$.} The linear dispersion relation for electrostatic waves propagating at $30^{\circ}$ to  an external magnetic field, with $\kappa=3$, $\Omega_{c}=0.5$.
 The upper (three) curves and the lower (three) curves depict $\omega_C$ and $\omega_A$, respectively.
 Different values of $\beta$ have been considered:
 $\beta=0.5$ (solid curve); $\beta=1$  (dashed curve); and $\beta=1.5$ (dot-dashed curve).}
 \label{Fig.3}
\end{figure}

\emph{Effect of cold-to-hot electron density ratio.} The dependence of the frequency on the cold-to-hot  electron concentration ratio  is depicted in Fig.~\ref{Fig.3}.
The frequency for both slow and fast modes increases with $\beta$, since the latter increases the inertia and decreases the restoring force underpinning the oscillations.

 \begin{figure}[!h]
 \begin{center}
 \includegraphics[width=8cm]{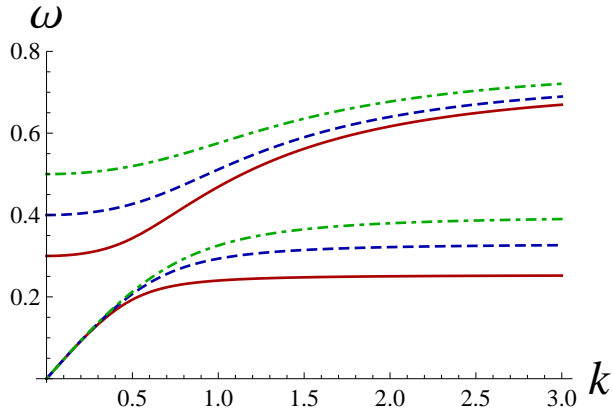}
 \end{center}
 \caption{(Color online) \emph{Effect of magnetic field strength}. The linear dispersion relation is depicted for electrostatic waves propagating at $30^{\circ}$ with respect to the external magnetic field, for $\kappa=3$, $\Omega_{c}=0.5$ and $\beta=0.5$.
 The upper (three) curves represent $\omega_C$, while the lower (three) curves represent $\omega_A$.
 Different values of $\Omega_{c}$ have been considered:
 $\Omega_{c}=0.3$ (solid curve); $\Omega_{c}=0.4$  (dashed curve); and $\Omega_{c}=0.5$ (dot-dashed curve).}
 \label{Fig.4}
\end{figure}

\emph{Magnetic field effect.} The quantitative influence of the magnetic field (expressed by $\Omega_c$) on the two modes is shown in Fig.~\ref{Fig.4}. A stronger external magnetic field increases the frequency gap between the two modes (slow and fast dispersion curves in the diagrams), as shown graphically in Fig. \ref{Fig.4}.
For normalized $k > 1$, a significant increase in the EA phase speed is seen to   occur for stronger magnetic field.

For perpendicular wave propagation, one may consider $\theta = \pi/2$ in Eq. (\ref{easagdis1}). The dispersion relation for the upper (cyclotron) mode thus becomes
 \begin{equation}
 \omega_{C}^{2}=\Omega_{c}^{2}+\tilde{\omega}^2\ , \label{easagdis3}
 \end{equation}
 or, in the long wavelength approximation, $\omega_{C}^{2}=\Omega_{c}^{2}+\frac{k^{2}\beta}{c_{\kappa}}$. On the other hand,  the lower (acoustic) mode vanishes in this case.

 Concluding, the cyclotron-like  mode tends to the wave frequency $\tilde{\omega}$ for parallel propagation, which is, in fact, the wave frequency for the EA modes in an unmagnetised plasma.
 On the other hand, the dispersion law recovers $\sqrt{\Omega_{c}^{2}+\tilde{\omega}^2}$ for perpendicular propagation. The frequency gap separating the two modes at $k=0$ is equal to $\Omega_c$,
but varies with $k$ for $k \geq 1$.

\section{Pseudopotential formalism for oblique electron-acoustic solitary waves \label{secSagdeev}}

We are now interested in large amplitude solitary structures. These may be analytically studied via the (so called Sagdeev-type) pseudopotential formalism, first introduced in Ref.~\cite{sagdeev1966}.
Anticipating stationary profile solitary waves moving at a constant speed, all state variables  are assumed to depend on a single variable, $\xi = \alpha x + \gamma z - M t$, where $M$ is the normalized
solitary wave propagation velocity scaled by a fixed sound speed (often referred to as the ``Mach number''), and  $\alpha$ and $\gamma$ are the direction cosines along the
$x$ and $z$ directions, respectively (thus, $\alpha^2 + \gamma^2 = 1$).
We note, for rigor, that the obliqueness (measured via the angle $\theta$ formed by the ambient magnetic field $B_0$ and the wave propagation vector $k$) must be small, in order to ensure a negligible current, which is a requirement for the electrostatic approximation to be valid for large deviations from equilibrium~\cite{FVerheest2009}. This is assumed throughout the nonlinear analysis below.

In order to solve for the anticipated solitary wave solutions, we integrate the continuity and momentum equation by considering vanishing boundary conditions at $\xi\rightarrow\infty$. The quasi-neutrality hypothesis (plasma approximation) is adopted  here, for analytical tractability, i.e., we no longer allow for space-charge effects through the use of Poisson's equation~(\ref{easagpois1}), but instead equate the perturbed charge densities through
Eq.~(\ref{easagpois2}). Accordingly, the normalized cold electron density is expressed as
\begin{equation}
n \simeq  \frac{1}{\beta}+1-\frac{1}{\beta}\biggr(1-\frac{\phi}{\kappa-\frac{3}{2}}\biggr)^{-\kappa+\frac{1}{2}} \ \, . \label{easagpois2}
\end{equation}
A tedious algebraic manipulation, described in Appendix A, leads us to an energy type integral for the electrostatic potential $\phi$, in the form:
\begin{equation}
\frac{1}{2} \left (\frac{d \phi}{d \xi} \right)^2  + \Psi(\phi;M,\kappa,\beta)=0 \, . \label{se10}
\end{equation}
This is formally analogous to a classical mechanical problem of motion in a potential field, Eq.~(\ref{se10})  representing the energy balance equation for a particle of unit mass located in a position $\phi$ at time $\xi$ and moving at speed $d\phi/d\xi$.
The potential energy is given by a  pseudopotential $\Psi(\phi; M,\kappa,\beta,\gamma)$ which can be expressed as
\begin{eqnarray}
\Psi(\phi;M,\kappa,\beta,\gamma)
= \Omega_{c}^{2}\biggr\{\frac{M^{2}(\kappa-1/2)}{\beta(\kappa-3/2)}\biggr[\frac{1}{\beta}+1-\frac{1}{\beta}\biggr(1-\frac{\phi}{\kappa-3/2}\biggr)^{-\kappa+1/2}\biggr]^{-3}\nonumber\\
\times \biggr(1-\frac{\phi}{\kappa-3/2}\biggr)^{-\kappa-1/2}-1\biggr\}^{-2} \times\biggr\{\frac{\phi}{\beta} + \frac{\gamma^{2} (1 + \beta) \phi}{
 M^{2} \beta^2} + \frac{\gamma^{2}(1+\beta)^{2}\phi^2}{2M^{2}\beta^{2}} \nonumber\\
+ \biggr[\frac{\gamma^{2}(1+\beta)}{M^{2}\beta^2}\, \frac{\kappa-3/2}{\kappa-5/2}\,(1-\phi) - \frac{\gamma^{2}+M^{2}\beta}{M^{2}\beta^2} \biggr] \biggr(1-\frac{\phi}{\kappa-3/2}\biggr)^{-\kappa+3/2}\nonumber\\
-\frac{\gamma^{2}(1+\beta)}{M^{2}\beta^2}\frac{\kappa-3/2}{\kappa-5/2}\biggr(1-\frac{\phi}{\kappa-3/2}\biggr)^{-\kappa+5/2}  +\frac{\gamma^2}{2M^{2}\beta^2}\biggr(1-\frac{\phi}{\kappa-3/2}\biggr)^{-2\kappa+3}\nonumber\\
+ \frac{M^{2}\beta +\gamma^2}{2(1+\beta)} \frac{1-\frac{\phi}{\kappa-3/2}}{\biggr[\biggr(1-\frac{\phi}{\kappa-3/2}\biggr)^{1/2}
 -(1+\beta)\biggr(1-\frac{\phi}{\kappa-3/2}\biggr)^{\kappa}\biggr]^2}\nonumber\\
+\frac{M^{2}\beta +\gamma^{2}(1+\beta)}{2(1+\beta)^2} \frac{\biggr(1-\frac{\phi}{\kappa-3/2}\biggr)
 -2(1+\beta)\biggr(1-\frac{\phi}{\kappa-3/2}\biggr)^{\kappa+1/2}}{\biggr[\biggr(1-\frac{\phi}{\kappa-3/2}\biggr)^{1/2}
 -(1+\beta)\biggr(1-\frac{\phi}{\kappa-3/2}\biggr)^{\kappa}\biggr]^2}\nonumber\\
- \frac{[-3-4(\kappa-2)\kappa]\phi+(4\kappa-2)\phi^{2}+4(1+\beta)(\kappa-3/2)^{2}
 \biggr(1-\frac{\phi}{\kappa-3/2}\biggr)^{\kappa+1/2}}{\biggr(1-\frac{\phi}{\kappa-3/2}\biggr)-
 (1+\beta)\biggr(1-\frac{\phi}{\kappa-3/2}\biggr)^{\kappa+1/2}}\nonumber\\
\times \frac{\gamma^2}{4(\kappa-3/2)^2} + \frac{M^2}{2(1+\beta)^2} + \frac{\gamma^2}{2M^{2}\beta^2}-\frac{\gamma^{2}\beta(2+\beta)-1-\beta}{\beta(1+\beta)}\biggr\}\ . \  \label{se11}
\end{eqnarray}

We may now study the nonlinear dynamics of EAWs by investigating  Eq.~(\ref{se11}) in terms of different plasma parameters within our model.

For the sake of reference, we may add that for a Maxwellian plasma, i.e., considering the limit $\kappa\rightarrow\infty$, expression~(\ref{se11}) reduces to
\begin{eqnarray}
\Psi_{Max}(\phi;M,\beta,\gamma)&=&\frac{\Omega_{c}^{2}}{2}\left\{1+\frac{e^{\phi}M^{2}\beta^{2}}{(e^{\phi}-1-\beta)^3}\right\}^{-2}\,\times
\, \biggr\{\frac{M^2}{(1+\beta)^2}+\frac{M^{2}\beta^{2}}{(e^{\phi}-1-\beta)^2}\nonumber\\
&+&\frac{M^{2}\beta(2+\beta)-\beta(1+2\beta)\gamma^2}{(1+\beta)^2} + \frac{\gamma^{2}\phi^2}{M^2} + \frac{\gamma^{2}(1-e^{\phi}+\phi)^2}{M^{2}\beta^2}\nonumber\\
&+&\gamma^{2}\biggr[2\phi\biggr(1+\frac{1}{e^{\phi}-1-\beta}\biggr)-\frac{2+3\beta}{(1+\beta)^2}\biggr] + \frac{2\beta[M^{2}-\gamma^{2}(1-\phi)]}{e^{\phi}-1-\beta}\nonumber\\
&-&\frac{2(e^{\phi}-1-\phi)(M^{2}+\gamma^{2}\phi)}{M^{2}\beta}\biggr\}\ . \label{sagMax}
 \end{eqnarray}
We may remark, not without surprise, that if we set  $\Omega_{c}=0$, expression~(\ref{se11})  reduces to nil, viz., $\Psi(\phi) = 0$.
This fact shows that our model fails to describe solitary waves in an unmagnetized plasma, a limitation imposed by the fact of having adopted the quasineutrality hypothesis in the algebra.

\subsection{Approximate analysis for small amplitude solitary waves}

So as to obtain some insight into the behaviour of the solitary waves, it is useful to consider first a small amplitude approximation.
In this limit ($\phi \ll 1$), one may expand the pseudopotential $\Psi(\phi)$
in Eq.~(\ref{se11}) as a McLaurin series near $\phi \approx 0$.
A brief algebraic manipulation then leads to
\begin{eqnarray}
\Psi_1(\phi;M,\kappa,\beta,\gamma) \approx \frac{\Omega_{c}^2}{2}\,\frac{M^{2}- \gamma^{2}\,M_2^{2}}{M^{2}\biggr(M^{2}-M_2^{2}\biggr)} \, \phi^2 \qquad \qquad \qquad \qquad \nonumber\\
\qquad \qquad +\frac{\Omega_{c}^{2}}{6\bigr(M_2^2-M^{2} \bigr)^2}\biggr\{15\gamma^{2} + \frac{\beta(4\gamma^{2}-1)(2\kappa+1)}{2\kappa-1}
\nonumber\\ \qquad \qquad \qquad \qquad \qquad \qquad 
- \frac{3 \gamma^{2}M_2^2}{M^{2}} -\frac{3 M^2}{M_2^2}\frac{\beta-4+2(4+\beta)\kappa}{2\kappa-1}\biggr\} \, \phi^{3}\ ,   \label{approxPotential1}
\end{eqnarray}
where we have defined the characteristic speed  $M_2 = [\beta (2 \kappa-3)/(2 \kappa-1)]^{1/2}$, which we recognize as the phase speed in an unmagnetized plasma, see in Eq.~(\ref{vph}).
Eq.~(\ref{approxPotential1}) describes weakly nonlinear waves at a speed $M$ (albeit within a range, to be determined below).

Taking the small amplitude analysis an order higher, we consider the weakly superacoustic range, traditionally described by a Korteweg-de Vries (KdV) type approach~\cite{Verheest1999}.
The true acoustic speed, which  provides the  lower bound of the Mach number range of permitted values, may be found by considering the roots, for a pseudopotential $\Psi(\phi)$, of the equation $\Psi''(0) =0$.
Using the above pseudopotential (Eq.~[\ref{approxPotential1}]), one finds  from $\Psi_1''(0) =0$ that the acoustic speed  for oblique (linear) electrostatic waves is given, in our case, by
$M_1 = |\gamma| M_2 = M_2 \cos\theta$: see Eq.~(\ref{vph}).

Upon setting $M \simeq M_1 + \tilde v = |\gamma| M_2 + \tilde v$, where $\tilde v \ll M_1$, and linearizing in
$\tilde v$,  expression~(\ref{approxPotential1}) is cast in the form:
\begin{eqnarray}
\Psi_2(\phi; \tilde v,\kappa,\beta,\gamma)    &\approx&  \frac{\Omega_{c}^2}{\gamma M_{2}^{3}(1-\gamma^2)}\,\tilde v \, \phi^2 + \frac{\Omega_{c}^2}{6 M_{2}^{4}(1-\gamma^2)^2}\biggr\{\frac{\beta(4\gamma^2-1)(2\kappa+1)}{2\kappa-1}\nonumber\\
&+&15\,\gamma^{2} - 3 - 3\gamma^{2}\,\frac{\beta-4+2(4+\beta)\kappa}{2\kappa-1}\biggr\} \, \phi^{3}  \label{approxPotential2}  \ .
\end{eqnarray}

The smallest non-zero root that annuls the pseudopotential yields the amplitude of the solitary waves, $\phi_m$, i.e.,  $\Psi_2 (\phi_m;\tilde v,\kappa,\beta,\gamma) = 0$ provides the amplitude. This leads to
\begin{equation}
|\phi_{m}| \approx f(\kappa,\beta,\gamma)\,\tilde{v}\, , \label{smallampPhi1}
\end{equation}
where
\begin{eqnarray}
f(\kappa,\beta,\gamma) \approx \frac{6\,M_{2}(1-\gamma^{2})}{\gamma} \nonumber \\  \qquad \qquad \times \biggr[15\,\gamma^{2} - 3+\frac{\beta(4\gamma^2-1)(2\kappa+1)}{2\kappa-1}- 3\gamma^{2}\,\frac{\beta-4+2(4+\beta)\kappa}{2\kappa-1}\biggr]^{-1}\, , \label{smallampPhi2}
\end{eqnarray}
and $\tilde{v}\simeq M - M_{1}$.
It is clear from Eq. (\ref{smallampPhi1}) that
$|\phi_{m}|\propto f(\kappa,\beta,\gamma)$ [or $|\phi_{m}|\propto F(\kappa,\beta,\gamma)$], where the proportionality constant takes the form $M-M_1$ or $\bigr[M/M_{1} - 1$ with $F(\kappa,\beta,\gamma)= |\gamma| \sqrt{\frac{\beta(2\kappa-3)}{2\kappa-1}}\,f(\kappa,\beta,\gamma)\bigr]$.

We shall return to this approximate solution for small amplitude solitons later in the paper.

\section{Existence domain for EA solitons \label{speeddomain}}

We shall now investigate the existence conditions and domains for arbitrary amplitude electron-acoustic solitary structures, based on the full nonlinear pseudopotential, Eq.~(\ref{se11}).

First, let us point out that the pseudopotential given by our plasma model in Eq.~(\ref{se11}) satisfies indeed the condition $\Psi(\phi=0)=\frac{d \Psi}{d\phi}|_{\phi=0} = 0$, since $\phi=0$ defines the equilibrium state. Furthermore, we recall that the maximum (absolute) value attained by the electrostatic potential $\phi_m$ is given by the root of the pseudopotential $\Psi(\phi)$ at $\phi_{m}\neq 0$.

The convexity requirement
$\frac{d^{2}\Psi}{d\phi^2}|_{\phi=0} \le 0$ (also referred to as the soliton existence condition~\cite{Verheest2008}) must be imposed, in order for the origin to correspond to a local maximum of the pseudopotential function $\Psi$. Expression~(\ref{se11}) thus leads to the analytical requirement
\begin{equation}
\frac{d^{2}\Psi}{d\phi^2}\biggr|_{\phi=0}=
\Omega_{c}^{2}\,\frac{M^{2}-\gamma^{2}\beta/c_{\kappa}}{M^{2}(M^{2}-\beta/c_{\kappa}\,)} \le  0\, . \label{mach1}
\end{equation}

For the sake of reference and comparison, we may adopt the notation of Ref.~\cite{Verheest2007} (see Eq. $5$ therein) by writing the convexity requirement given by Eq.~(\ref{mach1}) as
\begin{equation}
f(M):=\frac{(1-\gamma^{2})\Omega_{c}^{2}}{\beta/c_{\kappa}-M^2} +\frac{\gamma^{2}\,\Omega_{c}^{2}}{0-M^2}> 0 \ . \label{mach2}
\end{equation}
Note that Eqs.~(\ref{mach1}) and~(\ref{mach2}) are only valid for $\gamma\neq 1$; in other words, they are not satisfied for parallel propagation, $\theta = 0$.

The convexity requirement~(\ref{mach1}) imposes that the soliton speed $M$ value should lie between a threshold $M_1$ and an upper bound $M_2$, as
\begin{equation}
M_{1}=|\gamma|\sqrt{\beta/c_{\kappa}}\ , \qquad M_{2}=\sqrt{\beta/c_{\kappa}}\ . \label{machcrt1}
\end{equation}
The two limiting values are simply related via a factor $\gamma = \cos \theta$, related to the propagation direction (with respect to  the external magnetic field). Interestingly, the upper limit $M_2$ is essentially the phase speed of electron-acoustic waves in an unmagnetized plasma (see discussion in section~\ref{linearanalysis}) and does not depend on the direction of propagation. On the other hand,
the threshold $M_1$ is the real sound speed for obliquely propagating (linear) waves --  cf.~(\ref{EAphasemag}) above --  thus naturally, solitons occur above this limit.

Summarizing, EA solitons exist in the velocity range
\begin{equation}
M_{1}<M<M_{2}\ , \nonumber
\end{equation}
i.e.,
\begin{equation}
1<\frac{M}{M_{1}}<\gamma^{-1}\ . \label{mach3}
\end{equation}
One notes that, for any values of $\kappa$ and $\beta$, the effect of increasing the value of $\gamma$ (i.e., of aligning the wave vector to the direction of the magnetic field) is to shrink the existence region for EA solitary structures. We add, for rigor, that the existence domain vanishes in the limit of parallel propagation (see the definitions of $M_1$ and $M_2$, which then coincide for $\gamma = 1$), and the analytical model breaks down.

\begin{figure}[!h]
\begin{center}
\includegraphics[width=7cm]{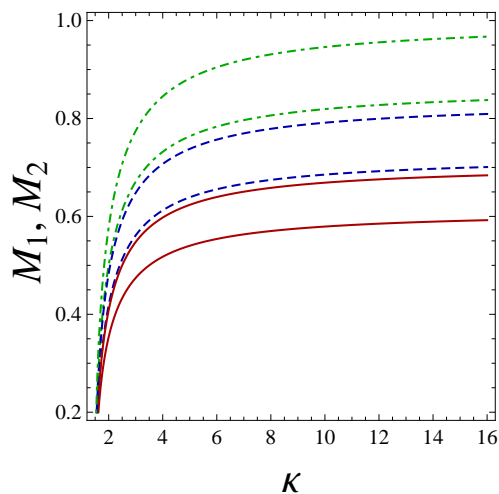}
\end{center}
\caption{(Color online) Variation of the lower threshold $M_{1}$ (lower curves) and the upper limit $M_2$ (upper curves) with the  superthermality parameter $\kappa$ for $30^{\circ}$ obliquely propagating waves. The curves correspond to $\beta=0.5$ for the solid (red) curves; $\beta=0.7$ for the dashed (blue) curves; and $\beta=1$ for the dot-dashed (green) curves.}
\label{Fig.5}
\end{figure}

The soliton existence region $[M_{1},\,M_{2}]$ is depicted in Fig.~\ref{Fig.5} for a representative choice of the parameter values, for different values of the cold electron concentration $\beta$.
We see that $M$
($\in \, [M_{1},\,M_{2}]$) varies significantly for $\kappa < 6$, whereas the interval $[M_{1},\,M_{2}]$ remains practically constant for $\kappa > 6$.
This is due to the fact that the acoustic limit $M_1$ decreases rapidly as $\kappa\to 3/2$. As expected, the plasma behaves as effectively Maxwellian for large values of $\kappa$ (practically, above $\simeq 10$).

Interestingly, the existence regions predicted for two different values of $\beta$ may not overlap at all (observe
Fig.~\ref{Fig.5} to see this, for instance), showing that solitary waves propagating at a given speed may exist for one value of $\beta$ but not for another.

We note here that there are a number of analogies between the oblique propagation of electron-acoustic and ion-acoustic solitary waves. Formally, the accessible range of Mach numbers naturally takes the same mathematical form as is the case for obliquely propagating ion-acoustic solitary waves~\cite{sultana2010}.
At the same time, it is clear that the  actual velocities would be very different, as the normalizing speed here is the hot electron thermal speed, whereas in the analogous case it is the much lower ion-acoustic speed.

We conclude this section by pointing out that the velocity interval (region between $M_1$ and $M_2$) for the existence of EA solitary waves leads to $M_{2}-M_{1} = (1-\cos \theta)\sqrt{\beta/c_{\kappa}}$.
The velocity interval is therefore expected to be reduced for lower values of  $\beta$, obliquity or superthermality parameter $\kappa$.
The upper limit $M_2$ does not depend on the direction of propagation.

\section{Electron-acoustic soliton characteristics \label{EAcharac}}

We are interested in investigating the features of large-amplitude electron-acoustic solitons [corresponding to solutions of Eq.~(\ref{se10}), obtained numerically], in fact focusing on their  dependence on various plasma parameters (superthermality parameter $\kappa$, cold electron concentration via $\beta$,  propagation direction via $\theta$). Numerical integration of Eq.~(\ref{se10}) provides us with the information about the electrostatic potential $\phi$ and also the corresponding electric field excitation $E\,(=-\nabla\phi)$.
We restrict  ourselves to a region where the linear EA wave may be sufficiently weakly damped to be observable, i.e., a region of cold-to-hot electron density ratio $\beta$ supported by Eq.~(\ref{relnbetakappa}) in order to ensure the validity of the fluid model, as well as respecting the limits imposed by Landau damping on the linear wave.

\begin{figure}[!h]
\begin{center}
\includegraphics[width=7.2cm]{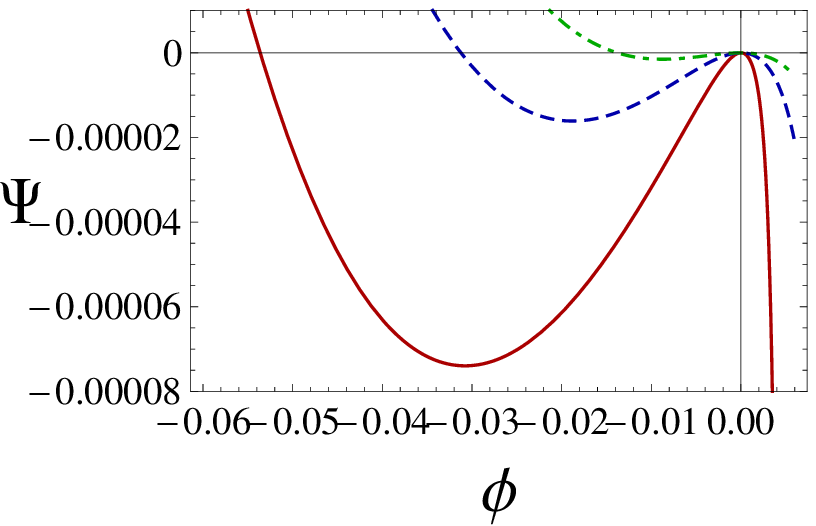}\hspace{0.5cm}
\includegraphics[width=6.9cm]{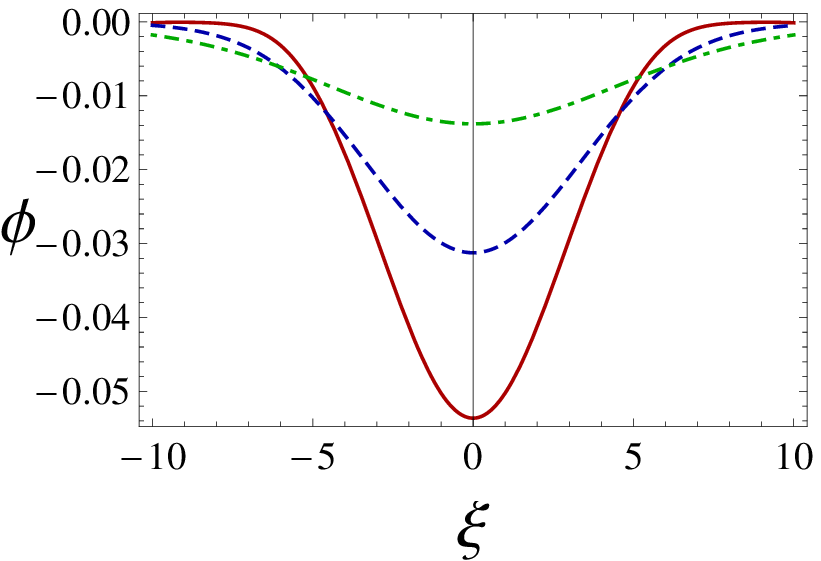}
\vspace{0.3cm}
\centerline{\qquad\Large{(a)}  \qquad \qquad \qquad \ \ \ \qquad \qquad   \Large{(b)}}
\end{center}
\caption{(Color online) (a) Variation of the pseudopotential $\Psi(\phi)$  with $\phi$ for different values of $\kappa$, and $\Omega_{c}=0.5$, $\beta=0.5$, $M=0.53$, $\theta=30^{\circ}$.
From top to bottom: dot-dashed curve for $\kappa=4$;  dashed curve  for $\kappa=3.5$ and solid curve for $\kappa=3$. (b) The corresponding electrostatic potential perturbations, obtained numerically.}
 \label{Fig.6}
\end{figure}

\textit{Superthermality effect (via $\kappa$).}
We focus on a low range of values for $\kappa$, to investigate the effect of superthermality on EA solitary structures. The soliton speed $M$ is kept constant here, within the accessible range of values.
Fig.~\ref{Fig.6} shows the effect of excess superthermality on the soliton characteristics, for a fixed speed value ($M = 0.53$) representative of the accessible velocity region (cf. Fig.~\ref{Fig.5} for $\beta = 0.5$).
One sees in Fig.~\ref{Fig.6}(a)  that  both the root and the depth of the pseudopotential well increase with decreasing $\kappa$.
This means that both the amplitude and the profile sharpness (maximum electric field) of the negative potential EA solitary waves (Fig.~\ref{Fig.6} [b]) increase  with increasing superthermality.
This is qualitatively reminiscent of the results for IA solitons in Ref.~\cite{sultana2010} (for magnetized plasma) and~\cite{saini2009} (for unmagnetized plasma).

The threshold above which the solitary waves may occur, namely the true acoustic speed $M_1$, has earlier been identified as the phase speed [cf.~Eq.~(\ref{EAphasemag}) above] of obliquely propagating EA excitations; this depends on both $\kappa$ and $\beta$.
One thus expects that the true Mach number $M/M_{1}$ (ratio of the pulse propagation speed to the true acoustic speed for the plasma composition under consideration), which differs for different parameter values (i.e., for different $\kappa$ and $\beta$) even for a fixed soliton propagation speed $M$, will be a meaningful physical quantity, whose effect on the soliton characteristics should be investigated.
Furthermore, it is known~\cite{Verheest2010a,Verheest2010b} that the pseudopotential function satisfies $\partial\Psi/\partial M < 0$, hence $\partial\phi_{m}/\partial M > 0$~\cite{Verheest2010a,Verheest2010b} (here $\phi_m$ is the maximum amplitude of the solitary excitations) for barotropic plasma fluids, suggesting that the soliton amplitude is an increasing function of $M/M_{1}$.
In addition, the amplitude of the the Korteweg-de~Vries (KdV)  solution (valid for small potential amplitudes)~\cite{FVMAHPS}   has a linear dependence on the supersonic velocity increment [i.e.,  $\propto (M - M_1)$, or $\propto M_1 (M/M_1 - 1)$, having adopted an appropriate scaling], as shown in Eq.~(\ref{smallampPhi1}).

\begin{figure}[!h]
 \begin{center}
\includegraphics[width=7.2cm]{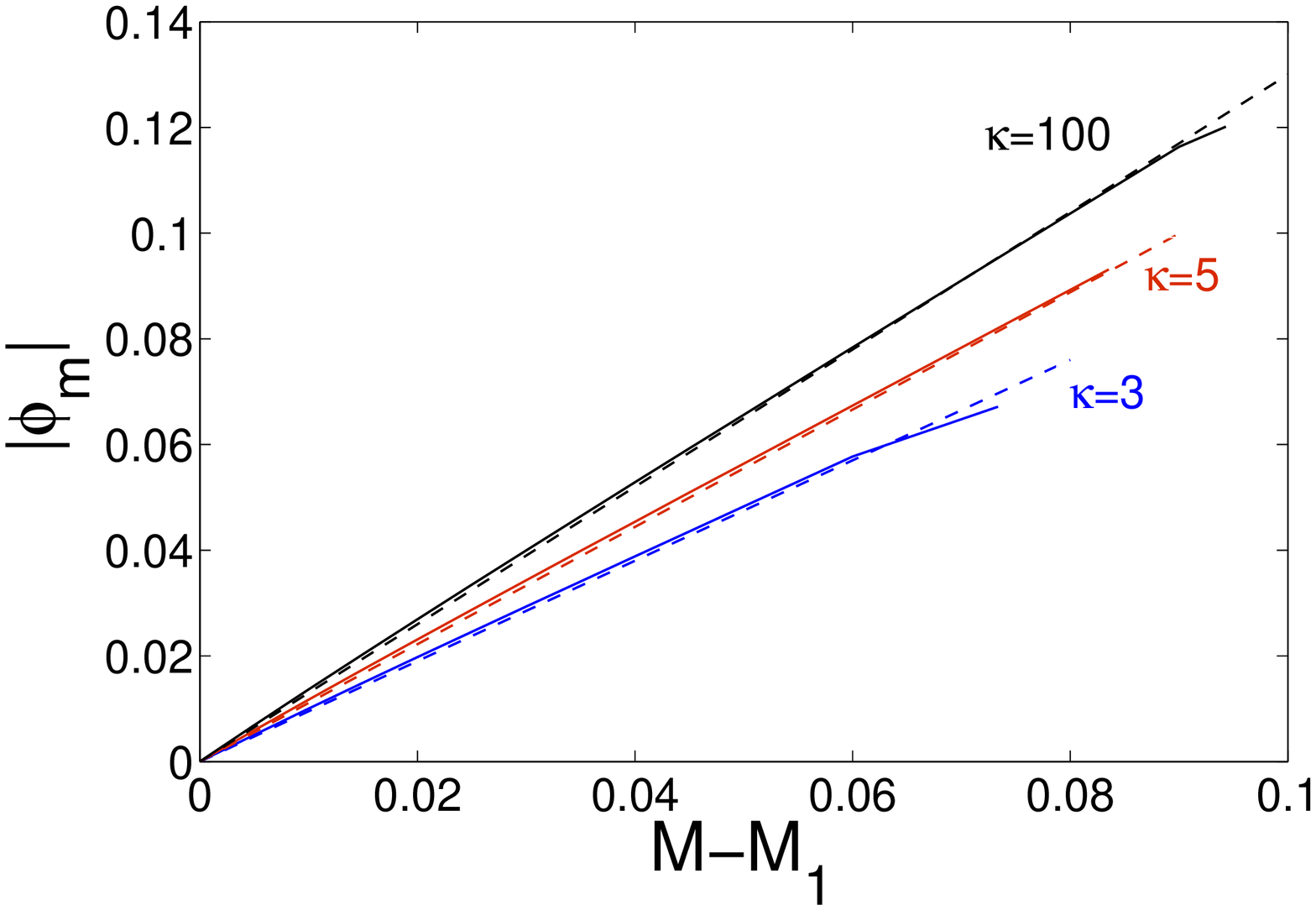}\hspace{0.5cm}
\includegraphics[width=7.8cm]{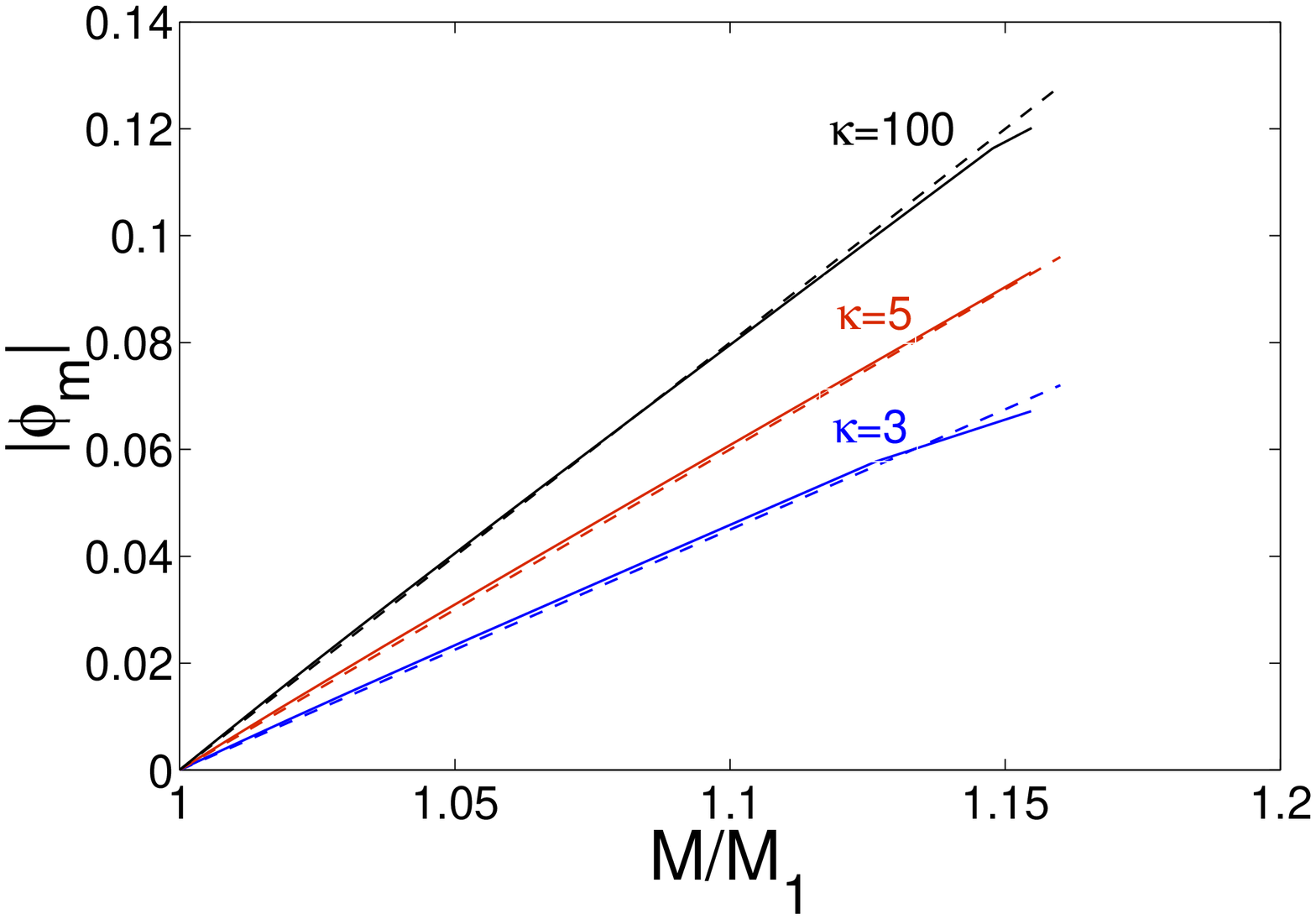}
\vspace{0.3cm}
\centerline{\qquad\Large{(a)}  \qquad \qquad \qquad \ \ \ \qquad \qquad   \Large{(b)}}
\end{center}
 \caption{(Color online) Variation of the pulse amplitude $|\phi_{m}|$ (absolute value) with (a) the super-acoustic excess speed, $M-M_1$, and (b) the true Mach number $M/M_1$, for different superthermality parameter values $\kappa$, as shown on the figures. The other parameters are fixed at $\beta=0.5,\,\Omega_{c}=0.5$ and $\theta=30^{\circ}$. Note that the curves have been  truncated at  the maximum accessible values of $M/M_{1}$, namely at $=M_{2}/M_{1}=1.15466,\,1.15469,\,1.1547$ for $\kappa=3,\,5,\,100$, respectively. The dashed lines represent the best fits to obtain values for the slopes.}
  \label{Fig.7}
\end{figure}

We have numerically obtained and investigated the variation of the  soliton amplitude $\phi_m$ as a function of  $M-M_1$, for different values of the relevant parameters, in particular, the superthermality parameter, $\kappa$, and the cold-to-hot electron density ratio, $\beta$.
The dependence of the acoustic speed $M_1$ on $\beta,\,\kappa$ and $\gamma$ has been taken into account, for a given plasma composition.
Fig.~\ref{Fig.7} shows the soliton amplitude $\phi_m$ against  $M-M_1$ and against $M/M_1$ for $30^{\circ}$ propagation (the true Mach number lies in the range $[1,\, 1.1547]$; see definition in Eq.~(\ref{mach3})), using the values $\beta=0.5$ and $\Omega_{c}=0.5$.
Both graphs show  practically linear variation of the amplitude,
but with the slope being significantly $\kappa$-dependent, particularly for small $\kappa$.
The dotted lines represent best fits, and in Fig.~\ref{Fig.7}(a) yield fitted slopes of $0.95, 1.11$ and $1.30$, for the three values of $\kappa = 3, \ 5, \ \rm{and} \ 100$, respectively, the last value clearly describing a quasi-Maxwellian case.
As the amplitudes are relatively small, one may expect that the KdV approximate solution should provide a reasonable description of the figures.
Indeed, substitution of the parameter values in Eqs.~(\ref{smallampPhi1},\ref{smallampPhi2}) leads to slopes of $1.02, 1.19$ and $1.39$, respectively.
Similar results are found when fitting to the slopes in Fig.~\ref{Fig.7}(b).
We note that the KdV expressions over-estimate the observed slopes by a few percent.
This is understandable, because the actual curves slowly flatten out and slip below a true linear form as $M-M_1$, and hence $|\phi_m|$, increase towards the limit of applicability of the small amplitude approximation.
This leads, for instance, to the observation that, for ($M-M_1) < 0.06$, the fitted (dashed) curve lies marginally below the calculated roots.

At first sight, the results presented in the Figs.~\ref{Fig.6} and~\ref{Fig.7} appear to be contradictory. From Fig.~\ref{Fig.6}, one might conclude that lower $\kappa$ implies stronger solitons, for fixed speed $M$. On the other hand, Fig.~\ref{Fig.7} suggests exactly the opposite trend, in that lower $\kappa$ implies lower amplitude $|\phi_m|$ for given $M/M_1$. The explanation lies in the fact that the former figure relates to a fixed speed $M$,  which implies a variable excess  speed (or the true Mach number in Fig.~\ref{Fig.7}) relative to the the true acoustic speed for the plasma composition under consideration. Thus the changes observed as $\kappa$ is varied relate strongly to the effect of $\kappa$ on $M_1$. On the other hand, in the latter figure, that effect is eliminated and one is left with the changes in amplitude that follow on the true variation from the sound speed (or the true Mach number) alone. Therefore, while Fig.~\ref{Fig.6} investigates the soliton characteristics in terms of a constant speed scale, Fig.~\ref{Fig.7} adopts a variable, $\kappa-$dependent yardstick, related to a fundamental physical unit, the true acoustic speed.

In qualitative terms, smaller values of $\kappa$ do indeed lead to more intense solitary waves (potential pulses), for given speed $M$, as Fig.~\ref{Fig.6} suggests; on the other hand,
the solitons are then ``less superacoustic,'' in loose formulation, and thus inevitably of smaller amplitude (and this, for given value of $M/M_1$).

\begin{figure}[!h]
\begin{center}
\includegraphics[width=8cm]{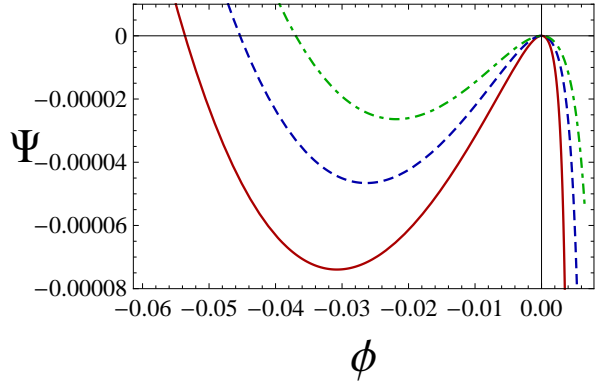}
\end{center}
\caption{(Color online) Variation of the pseudopotential $\Psi(\phi)$ for  different values of $\beta$, with $\phi$ for $\Omega_{c}=0.5$, $\kappa=3$, $M=0.53$, $\theta=30^{\circ}$.
From top to bottom: the dot-dashed curve for $\beta=0.54$; the dashed for $\beta=0.52$ and the solid for $\beta=0.50$.}
 \label{Fig.8}
\end{figure}

\textit{Effect of the cold electron concentration (via $\beta$).}
We next turn to an analogous study of the effects of the cold-to-hot-electron density ratio ($\beta$)  for a strongly superthermal plasma ($\kappa=3$).
We assume parameter values $\kappa = 3, M=0.53, \Omega_{c}=0.5, \theta =30^{\circ}$,
and vary  $\beta$ over  a narrow range.
One notes that, as $\beta = n_{0c}/n_{0h}$ is decreased, the relative inertia (the cold electron mass density governed through $n_{0c}$) is decreased, while the relative restoring force (the hot electron pressure governed through $n_{0h}$)   is increased.
These two effects both tend to increase the frequency of the electrostatic oscillations underlying the EA wave and soliton.
Hence, for the three cases illustrated in Fig.~\ref{Fig.8}, the true Mach number $M/M_{1}$ increases with decreasing $\beta$, as $M/M_1=1.075,\,1.095,\,1.117$ for $\beta=0.54,\,0.52,\,0.50$, respectively.
As a result, one would expect both the soliton amplitude and the maximum soliton electric field to increase.
In terms of the pseudopotential curves, these effects imply that one would expect an increase in both the position of the root and the well-depth of the pseudopotential.
We immediately recognise exactly these two effects in Fig.~\ref{Fig.8}.

\begin{figure}[!h]
 \begin{center}
\includegraphics[width=7.2cm]{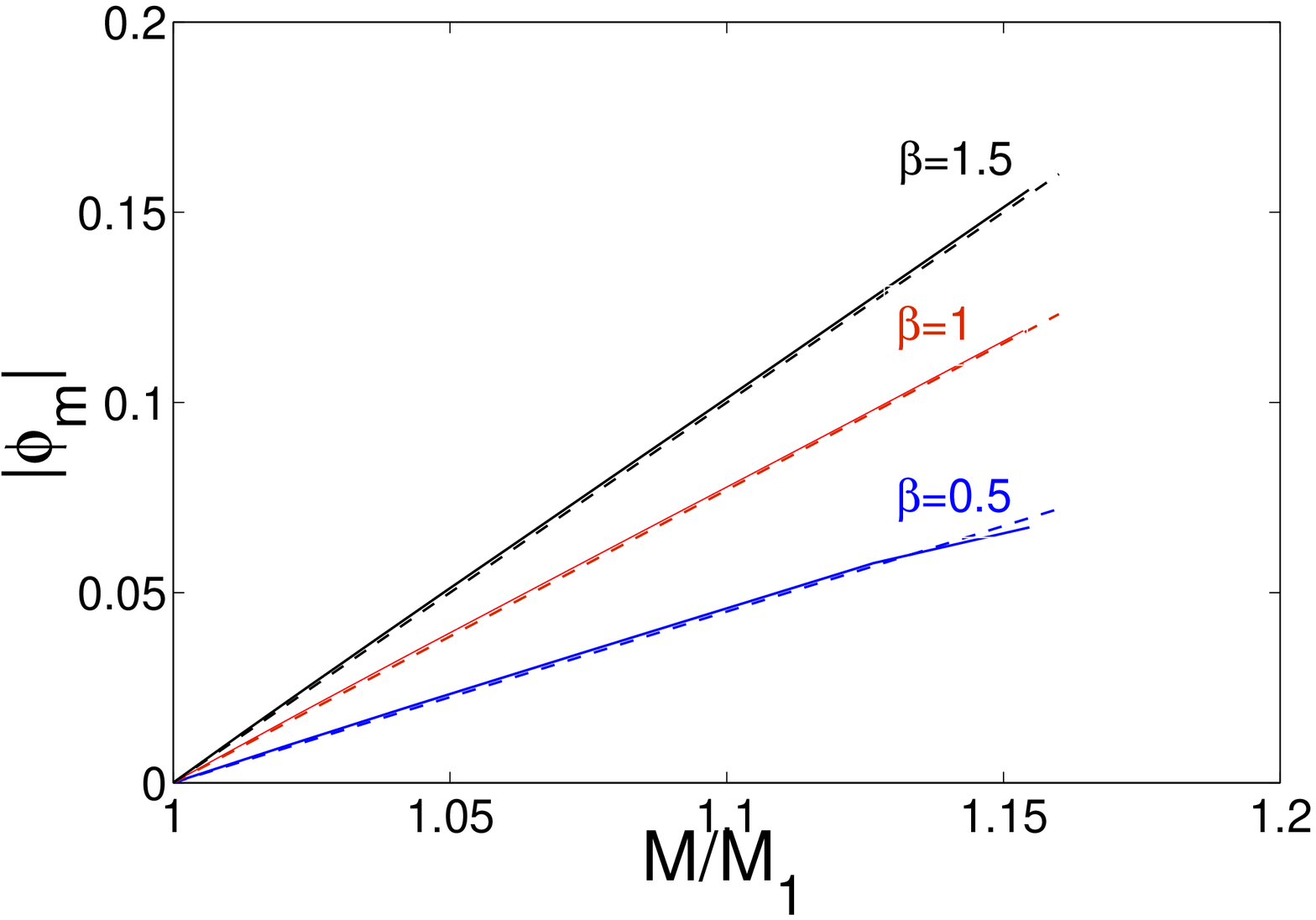}\hspace{0.5cm}
\includegraphics[width=7.8cm]{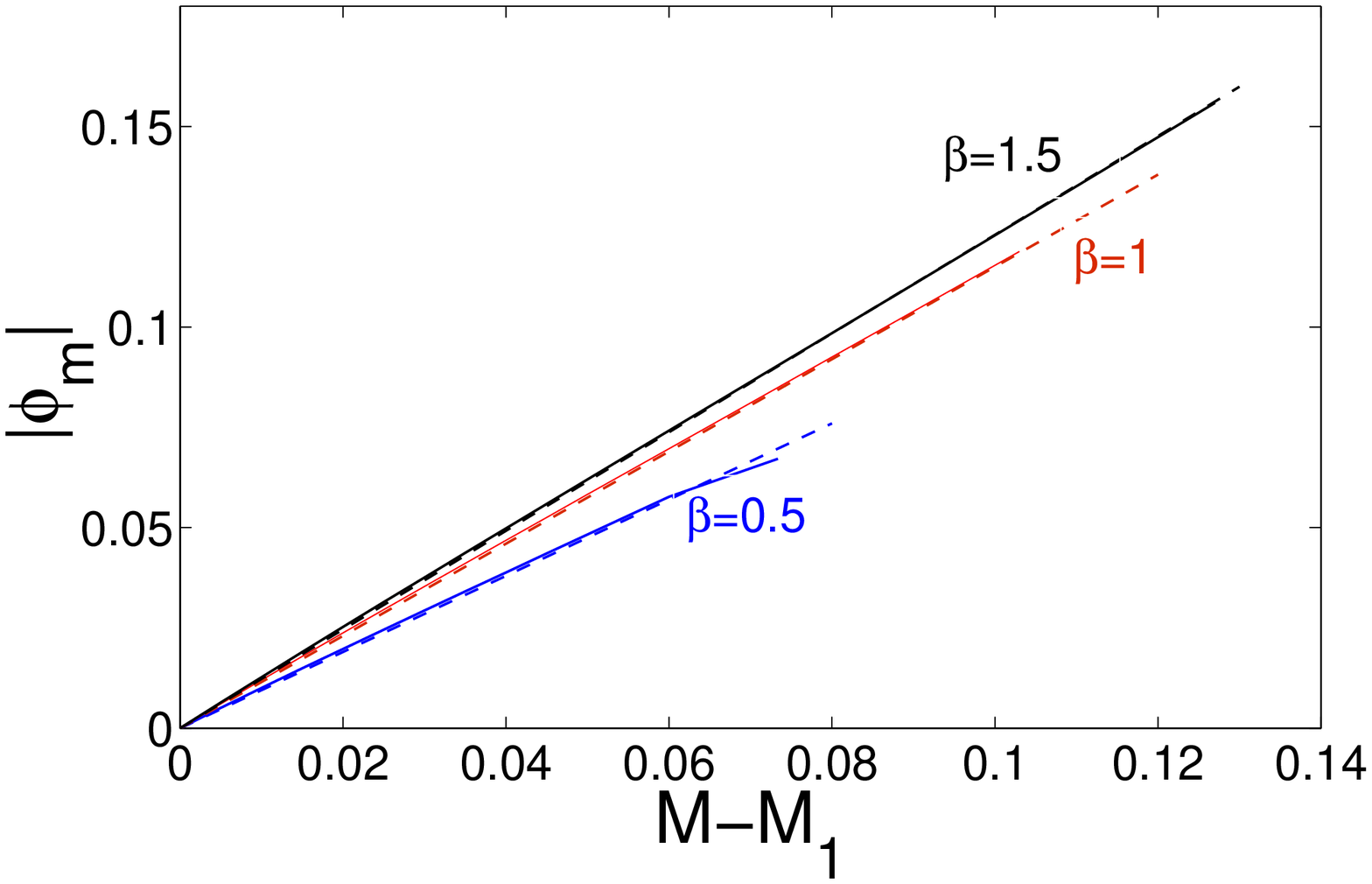}
\vspace{0.3cm}
\centerline{\qquad\Large{(a)}  \qquad \qquad \qquad \ \ \ \qquad \qquad   \Large{(b)}}
\end{center}
\caption{(Color online) Variation of the pulse amplitude $|\phi_{m}|$ (absolute value) with  (a) the true Mach number $M/M_1$; and (b) the super-acoustic excess in speed, $M-M_1$, for different cold-to-hot electron population ratio $\beta$. The other parameters are fixed at $\kappa=3,\,\Omega_{c}=0.5$ and $\theta=30^{\circ}$. Note that we have truncated the curves at $M/M_{1}=M_{2}/M_{1}=1.15466,\,1.15354,\,1.15458$ for $\beta=0.5,\,1,\,1.5$, respectively.
The dashed lines represent best linear fits to the data. }
 \label{Fig.9}
\end{figure}

In Fig.~\ref{Fig.9} we next present figures that are analogous to those of Fig.~\ref{Fig.7}, but varying $\beta$ rather than $\kappa$. Other parameter values are fixed at $\kappa = 3, \Omega_c = 0.5, \theta = 30^{\circ}$.
It will be seen that the values for $\beta$ that are used in Fig.~\ref{Fig.9} represent cases in which the cold and hot electron densities are equal, as well as cases in which the cold and the hot electron densities, respectively, dominate: the values used are $\beta = 0.5,\ 1.0,\ \rm{and} \ 1.5$.
As we have seen above, the sound speed (soliton velocity threshold), $M_1= [\beta (2\kappa-3)/(2 \kappa - 1)]^{1/2}$.
 One would thus expect behaviour with respect to increasing $\beta$ to be similar to that seen for increasing $\kappa$ in Fig.~(\ref{Fig.7}).

The variation of $\phi_m$ with $M/M_{1}$ and with $M-M_1$ is explored in Fig.~\ref{Fig.9}.
 The figure shows that with a decreasing  cold-to-hot electron density ratio, the true Mach number, and hence the maximum amplitude of the soliton increases  (see Fig.~\ref{Fig.9}).
 The best linear fit in Fig.~\ref{Fig.9}(b) yields values for the $\beta$-dependent slope of $0.95, 1.15,$ and $1.23$, respectively,  as $\beta$ is increased.
 These values are well reflected by the KdV solution, based on Eqs.~(\ref{smallampPhi1},\ref{smallampPhi2}), which yields slopes of $1.02, 1.2$ and $1.28$, respectively.
Again, one finds that the KdV solution over-estimates the fitted slopes by a few percent, as discussed above for the case in which $\kappa$ was varied.

\begin{figure}[!h]
\begin{center}
\includegraphics[width=7.1cm]{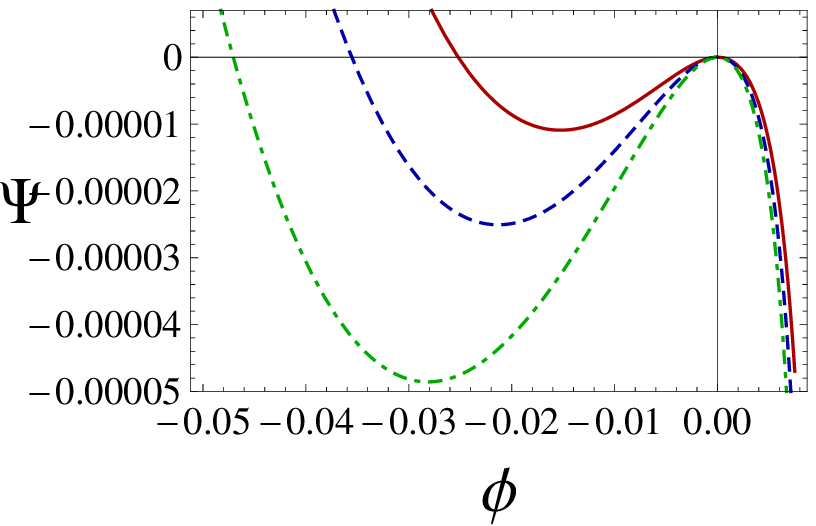}\hspace{0.5cm}
\includegraphics[width=6.8cm]{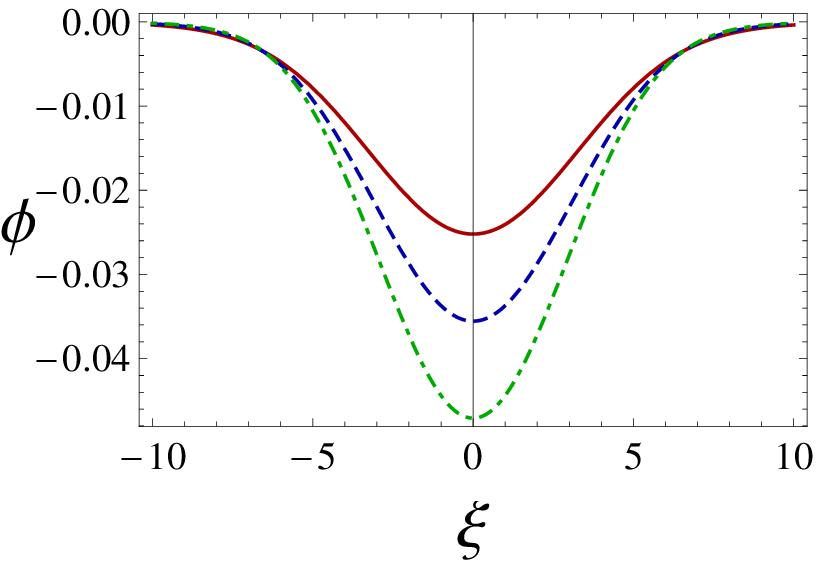}
\vspace{0.3cm}
\centerline{\qquad\Large{(a)}  \qquad \qquad \qquad \ \ \ \qquad \qquad   \Large{(b)}}
\end{center}
\caption{(Color online) Obliqueness effect, via $\theta$ (the angle between wave propagation direction and magnetic field) : (a) variation of $\Psi(\phi)$  with $\phi$ for varying $\theta$, and  $\Omega_{c}=0.5$, $\kappa=3$, $M=0.5$, $\beta=0.5$. From top to bottom:  solid curve for $30^{\circ}$; dashed curve for $32^{\circ}$ and dot-dashed curve for $34^{\circ}$. (b) The corresponding solutions for the electrostatic potential perturbation, obtained numerically.}
 \label{Fig.10}
\end{figure}

\paragraph{\textit{Obliqueness effect (via $\theta$ or $\gamma$).}}
We have analyzed the effect of obliqueness of propagation angle on electrostatic solitary waves, for given values of the other plasma parameters.
The Sagdeev potential, as well as the electrostatic potential (pulse), are amplified with  increasing
$\theta$, as shown in Fig.~\ref{Fig.10}.
In particular, both the root and the depth of the pseudopotential ($\Psi$) well are increased, thus leading to larger values of both the maximum derivative $d\phi/d\xi$ (maximum steepness of the soliton profile) and the soliton amplitude $|\phi_m|$, as shown in the plot.
This is intuitively expected.
We recall that the critical Mach number threshold $M_1$ decreases with increasing obliqueness as  $M_1 \propto \cos \theta$. Therefore, the difference $M-M_1$ for fixed $M$ increases with  $\theta$, resulting in both a deeper Sagdeev pseudopotential and to a larger root, and thus to stronger electrostatic excitations.

\begin{figure}[!h]
\begin{center}
\includegraphics[width=7.1cm]{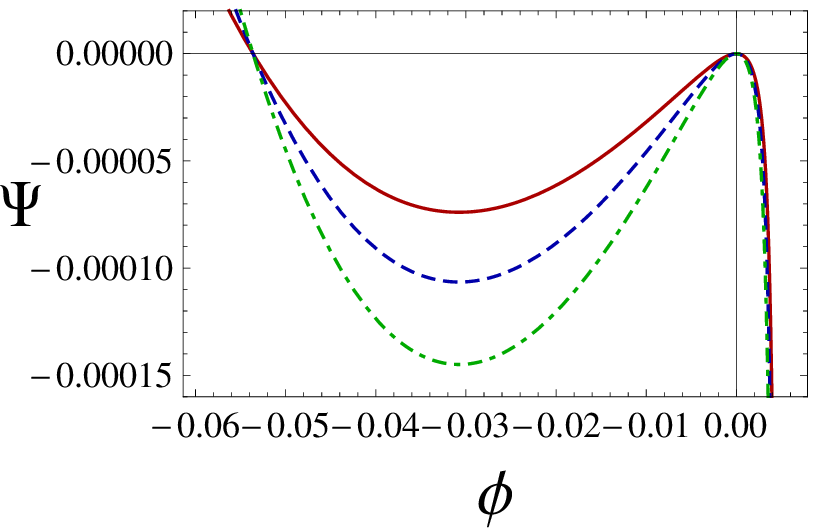}\hspace{0.5cm}
\includegraphics[width=6.8cm]{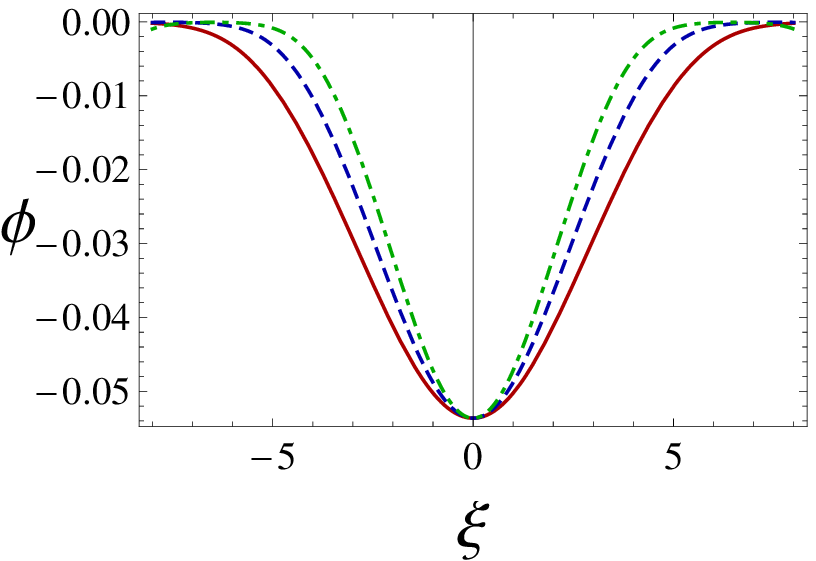}

\vspace{0.3cm}
\centerline{\qquad\Large{(a)}  \qquad \qquad \qquad \ \ \ \qquad \qquad   \Large{(b)}}
\end{center}
\caption{(Color online) Magnetic field effect (dependence on $\Omega_c$): (a) variation of $\Psi(\phi)$  with $\phi$ for $\kappa=3$, $M=0.53$, $\beta=0.5$, $\theta=30^{\circ}$.
From top to bottom: $\Omega_{c}=0.5$ (solid curve); $\Omega_{c}=0.6$ (dashed curve) and   $\Omega_{c}=0.7$ (dot-dashed curve). (b) The corresponding electrostatic pulse excitation, obtained numerically.}
 \label{Fig.11}
\end{figure}

\textit{Effect of Magnetic field (via $\Omega_c$).}
The existence domain $[M_{1},\,M_{2}]$ [discussed in Section~\ref{speeddomain}, see Eq.~(\ref{machcrt1})] for the existence of EA solitary excitations does not depend on the magnetic field $B_{0}$ (since it does not depend on the value of $\Omega_c$).  However, the depth of the pseudopotential well increases for larger $\Omega_{c}$, i.e., for stronger magnetic field $B_0$, as shown in Fig.~\ref{Fig.11}. The corresponding soliton solutions thus yield constant amplitude, but reduced soliton profile width (i.e., are more sharply peaked), for larger $\Omega_{c}$.

\section{Conclusions \label{EAsagconc}}

The characteristics of large amplitude obliquely propagating EA solitary waves have been investigated in a three component magnetized superthermal plasma.

We first briefly considered the linear modes and found two distinct types, namely, obliquely propagating magnetized electron-acoustic waves and electron cyclotron-like waves.
The frequency gap between the two modes decreases with obliqueness, i.e., with the angle between the wave propagation direction and the direction of the magnetic field.
On the other hand,  a simultaneous decrease of the frequency of both modes is observed for stronger superthermality (lower $\kappa$), i.e., for larger deviation from the Maxwellian, as well as  for an increase of the cold-to-hot electron population ratio, $\beta$.

The main emphasis of the paper is the study of nonlinear electrostatic solitary waves, using a pseudopotential method, adapted to treat obliqueness effects. For simplicity, we have adopted the quasineutrality hypothesis for our analytical study in the nonlinear part. We have determined and analyzed the velocity domain where solitary waves may occur, in the form of
a threshold $M_1$ and an upper bound $M_2$, both of which depend on $\kappa$ and on $\beta$.
Here, $M_1=M_2 \cos \theta$, where $\theta$ is the angle is the angle between the direction of propagation and the background magnetic field.
The relevant plasma configurational parameters (e.g., the plasma superthermality, the cold electron concentration, the propagation angle) are seen to significantly modify the allowed regions for the linear EAWs to propagate, along with the associated characteristics of solitary waves.

 We have shown that the true acoustic speed of electron-acoustic waves in a magnetized plasma decreases with stronger plasma superthermality, decreasing cold electron concentration and increasing obliqueness of propagation. This variation of the true acoustic speed with different plasma parameters therefore modifies the characteristic properties of solitary waves (e.g., amplitude and width).

For the sake of rigor, we have tested the validity of our oblique model in various special cases. Interestingly, the two limiting Mach-number values $(M_1, M_2)$ merge in the parallel propagation limit ($\theta \rightarrow 0$), indicating that the parallel wave propagation in magnetized plasma can not be described by this model. Similarly, the model breaks down in the vanishing-magnetic-field limit, which may be attributed to the plasma approximation, here adopted for analytical tractability.

Beyond the strict interest in the context of plasma waves, our study is of methodological interest, as it adopts an analytical approach not used before (with the sole exception of Ref.~\cite{sultana2010}, where this first appeared). In particular, the fluid-dynamical equations, which generally cannot be disentangled into a single pseudo-energy balance equation if oblique wave propagation in magnetized plasma is considered, were managed into a convenient form by assuming charge neutrality. There are a number of analytical issues thus raised.

 First of all, a comment is in order, regarding our choice of adopting the plasma approximation (neutrality hypothesis, NH). As a matter of fact, the question of whether or not the NH is appropriate for the modeling of electrostatic excitations appears to be as old as plasma modeling itself, and is discussed in early plasma textbooks~\cite{Braginskii, Chen}, yet on a qualitative, first-principles basis. As a rule of thumb, it was suggested that high-frequency waves should not be modeled by making use of the NH. Later literature has attempted to shed some light on the problem, with rigorous emphasis on localized nonlinear excitations (solitary waves), where the legitimate quantity to deal with is the phase speed, rather than the frequency (notice the discussion in Ref.~\cite{FV2007}). Furthermore, in the ``conventional'' version of the pseudopotential method, for magnetic-field-aligned solitary waves,  one encounters a second upper limit, related to the infinite compression point (or limit): this point cannot be accessed via the methodology adopted in this paper. This may be attributed to the physical and also analytical constraints implied by adopting the plasma approximation (neutrality hypothesis).

 Admittedly, our choice of adopting the NH was at first dictated by algebraic tractability (as the 2D fluid equations could not be disentangled into a simple evolution equation by retaining the full Poisson dynamics).
 The algebra leading from~(\ref{eo1})-(\ref{eo2}) via the NH to~(\ref{se10})-(\ref{se11}) is presented for the first time, to our knowledge. From a methodological point of view, therefore, we find that this study exhibits novelty and interest which should attract future attention and discussion, in other nonlinear plasma wave related problems.
Focusing on the particular area of electrostatic solitary waves, our results on the role of non-Maxwellian behaviour in plasmas include a number of clear predictions on soliton existence and stability, which may be tested, e.g., against space observations, and hopefully confirmed.

\ack
The UK Engineering and Physical Sciences Research Council (EPSRC) is acknowledged for the financial support under grant No. EP/D06337X/1.
One of the authors (M.A.H.) would like to acknowledges the National Research Foundation of South Africa (NRF) for financial support.
Any opinion, findings and conclusions or recommendations expressed in this material are those of the authors and therefore,  NRF does not accept any liability in regard thereto.

\section*{Appendix: Derivation of the Sagdeev-type pseudo-energy-balance equations}
Defining the moving coordinate $\xi=\alpha x + \gamma z - M t$ (recall that $\gamma = \cos\theta$)
one can write the normalized fluid equations (\ref{eo1}) and (\ref{eo2}) in the form
\begin{eqnarray}
-M\frac{d{n}}{d\xi} &+& \alpha \frac{d{(nu_x)}}{d \xi}+ \gamma \frac{d{(nu_z)}}{d \xi} = 0 \, , \label{ease1}\\
(-M &+& \alpha u_x+\gamma u_z) \frac{du_x}{d\xi}-\alpha \frac{d\phi}{d\xi}+\Omega_{c}u_y = 0 \, ,\  \label{ease2} \\
(-M &+& \alpha u_x+\gamma u_z) \frac{du_y}{d\xi}- \Omega_{c} u_x = 0 \, , \label{ease3} \\
(-M &+& \alpha u_x+\gamma u_z) \frac{du_z}{d\xi}-\gamma \frac{d\phi}{d\xi} = 0 \, . \label{ease4}
\end{eqnarray}
Anticipating stationary localized excitations (vanishing at infinity), we integrate Eqs.~(\ref{ease1}) and~(\ref{ease4}) using the boundary conditions $n\rightarrow1,u_{x,z}\rightarrow0, \ {\rm and} \ \phi\rightarrow0 \ {\rm at } \ \xi\rightarrow\pm\infty$, to obtain
\begin{eqnarray}
\alpha u_x\ &+& \gamma u_z = M\biggr(1-\frac{1}{n}\biggr), \label{eaden1} \\
u_z &=& -\frac{\gamma}{M} \biggr(\frac{1}{\beta}+\int{n\,d\phi}\biggr), \label{ease5} \\
u_x &=& \frac{M}{\alpha} \biggr(1-\frac{1}{n} \biggr)+\frac{\gamma^2}{M\alpha} \biggr(\frac{1}{\beta}+\int{n\,d\phi} \biggr) \, , \label{ease6}
\end{eqnarray}
where we have taken quasineutrality, i.e., Eq.~(\ref{easagpois2}), into account. Substituting Eq.~(\ref{eaden1}) into Eqs.~(\ref{ease2}) and~(\ref{ease3}), we can write
\beqa
-\frac{M}{n} \frac{d u_{x}}{d \xi}& - &\alpha \frac{d \phi}{d \xi} + \Omega_{c}u_{y}=0 \ , \label{ea1}\\
-\frac{M}{n} \frac{d u_{y}}{d \xi}& - &\Omega_{c}u_{x}=0 \ , \label{ea2}
\eeqa
Substituting for $u_x$ from Eq.~(\ref{ease6}), Eq.~(\ref{ea2}) can be written as
\beq
\frac{d u_y}{d\xi}=-\Omega_{c}\biggr[\frac{n}{\alpha}-\frac{1}{\alpha} + \frac{\gamma^2}{M^{2}\alpha}\biggr(\frac{n}{\beta}+n\int n d\phi\biggr)\biggr]\ . \label{ea3}
\eeq
Differentiating Eq.~(\ref{ea1}) with respect to $\xi$ and substituting the value of $u_x$ and $du_y/d\xi$, we get
\beq
\frac{3M^2}{n^4}\biggr(\frac{d n}{d\xi}\biggr)^{2}-\frac{M^2}{n^3}\frac{d^{2}n}{d\xi^2}-\frac{d^{2}\phi}{d\xi^2} = F(\phi)\ , \label{ea4}
\eeq
where
\begin{eqnarray}
F(\phi)& =& \Omega_{c}^{2}
\biggr[ -\frac{1}{\beta^2}\biggr(\beta+\frac{\gamma^2}{M^2}\biggr)\biggr(1-\frac{\phi}{\kappa-3/2}\biggr)^{-\kappa+1/2}
+\frac{1}{\beta}+\frac{\gamma^2}{M^{2}\beta^2}(1+\beta)\nonumber\\
&+&\frac{\gamma^2}{M^2}\biggr\{\frac{\phi}{\beta^2}+\frac{2\phi}{\beta}+\phi-\frac{1+\beta}{\beta^2}\biggr(1-\frac{\phi}{\kappa-3/2}\biggr)^{-\kappa+3/2}
\nonumber\\&-&\frac{\phi(1+\beta)}{\beta^2}\biggr(1-\frac{\phi}{\kappa-3/2}\biggr)^{-\kappa+1/2}
+\frac{1}{\beta^2}\biggr(1-\frac{\phi}{\kappa-3/2}\biggr)^{-2\kappa+2}\biggr\} \biggr] \, .  \label{se8}
\end{eqnarray}

We may now make use of  Eq.~(\ref{easagpois2}) in order to eliminate (the normalised density) $n$ in
Eq.~(\ref{ea4}), leading to an expression in terms of (the electrostatic potential) $\phi$ only,
\beqa
& &3M^{2}\frac{(\kappa-\frac{1}{2})^2}{\beta^2(\kappa-\frac{3}{2})^2} \,\biggr[\frac{1}{\beta}+1-\frac{1}{\beta}\biggr(1-\frac{\phi}{\kappa-\frac{3}{2}}\biggr)^{-\kappa+\frac{1}{2}}\biggr]^{-4}
\,\biggr(1-\frac{\phi}{\kappa-\frac{3}{2}}\biggr)^{-2\kappa-1}\biggr(\frac{d\phi}{d\xi}\biggr)^2\nonumber\\
&+& M^{2}\frac{\kappa^{2}-\frac{1}{4}}{\beta(\kappa-\frac{3}{2})^2} \,\biggr[\frac{1}{\beta}+1-\frac{1}{\beta}\biggr(1-\frac{\phi}{\kappa-\frac{3}{2}}\biggr)^{-\kappa+\frac{1}{2}}\biggr]^{-3}
\,\biggr(1-\frac{\phi}{\kappa-\frac{3}{2}}\biggr)^{-\kappa-\frac{3}{2}}\biggr(\frac{d\phi}{d\xi}\biggr)^2\nonumber\\
&+&\frac{M^{2}(\kappa-\frac{1}{2})}{\beta(\kappa-\frac{3}{2})}\,\biggr[\frac{1}{\beta}+1-\frac{1}{\beta}\biggr(1-\frac{\phi}{\kappa-\frac{3}{2}}\biggr)^{-\kappa+\frac{1}{2}}\biggr]^{-3}  \,\biggr(1-\frac{\phi}{\kappa-\frac{3}{2}}\biggr)^{-\kappa-\frac{1}{2}}\,\frac{d^{2}\phi}{d \xi^2}
-\frac{d^{2}\phi}{d \xi^2}=F(\phi)\nonumber
\eeqa

Differentiating twice with respect to  $\xi$, the above equation reduces to
\beqa
& &\frac{M^2}{2}\frac{d^2}{d \xi^2}\biggr[\frac{1}{\beta}+1-\frac{1}{\beta}\biggr(1-\frac{\phi}{\kappa-\frac{3}{2}}\biggr)^{-\kappa+\frac{1}{2}}\biggr]^{-2}-\frac{d^{2}\phi}{d \xi^2}=F(\phi)\nonumber\\
&\Rightarrow& \frac{d^2}{d \xi^2} \biggr(\frac{M^2}{2}\biggr[\frac{1}{\beta}+1-\frac{1}{\beta}\biggr(1-\frac{\phi}{\kappa-\frac{3}{2}}\biggr)^{-\kappa+\frac{1}{2}}\biggr]^{-2}-\phi\biggr)=F(\phi)\nonumber\\
&\Rightarrow& \frac{d^{2}S}{d \xi^2}=F(\phi) \ , \label{ea5}
\eeqa
where
\begin{equation}
S(\phi)=\frac{M^2}{2}\biggr[ \frac{1}{\beta}+1-\frac{1}{\beta}\biggr(1-\frac{\phi}{\kappa-\frac{3}{2}}\biggr)^{-\kappa+\frac{1}{2}}\biggr]^{-2} -\phi\ . \label{eas1}
\end{equation}
Differentiating $S$ with respect to $\xi$ and squaring, we write
\beq
\biggr(\frac{d S}{d \xi}\biggr)^2=\biggr[G(\phi)\biggr]^{2}\biggr(\frac{d \phi}{d\xi}\biggr)^2 \ ,\label{ea6}
\eeq
where
\begin{equation}
G(\phi)=\frac{M^{2}(\kappa-\frac{1}{2})}{\beta(\kappa-\frac{3}{2})}\biggr[\frac{1}{\beta}+1-\frac{1}{\beta}\biggr(1-\frac{\phi}{\kappa-\frac{3}{2}}\biggr)^{-\kappa+\frac{1}{2}}\biggr]^{-3}
\,\biggr(1-\frac{\phi}{\kappa-\frac{3}{2}}\biggr)^{-\kappa-\frac{1}{2}} - 1\ .
\end{equation}

Multiplying both sides of Eq.~(\ref{ea6}) by $d S/d \xi$, we can write
\beq
\frac{d}{d \xi}\biggr[\frac{1}{2}\biggr(G(\phi)\frac{d \phi}{d \xi}\biggr)^2\biggr]=F(\phi) G(\phi)\frac{d \phi}{d \xi} \ .\label{ea7}
\eeq

Integrating this equation with boundary conditions $\phi\rightarrow 0$ and $d \phi/d \xi\rightarrow 0$ at $\xi\rightarrow \pm\infty$, one can obtain an energy balance equation in the form
\begin{equation}
\frac{1}{2} \left (\frac{d \phi}{d \xi} \right)^2  + \Psi(\phi;M,\kappa,\beta)=0 \, , \label{energybalance}
\end{equation}
where $\Psi(\phi;M,\kappa,\beta)$ defines the Sagdeev-type pseudopotential of the system.

\end{document}